\documentclass{PoS}

\title{The Beautiful Physics of LHC Run 2}

\ShortTitle{Beautiful Run 2}

\author{\speaker{John Ellis}\thanks{Work supported by the European Research Council 
via the Advanced Investigator Grant 267352 and by STFC (UK) via the research grant ST/J002798/1.}\\
        Theoretical Particle Physics and Cosmology Group, Department of Physics, \\
King's College London, Strand, London WC2R 2LS, U.K; \\
Theory Division, Physics Department, CERN, CH 1211 Geneva 23, Switzerland\\
        E-mail: \email{John.Ellis@cern.ch}\\
        ~\\
      {\tt ~~~~~~~~~~KCL-PH-TH/2014-50, LCTS/2014-51, CERN-PH-TH/2014-248}  }

\abstract{Run 2 of the LHC offers some beautiful prospects for new physics, including flavour physics
as well as more detailed studies of the Higgs boson and searches for new physics beyond the
Standard Model (BSM). One of the possibilities for BSM physics is supersymmetry, and flavour physics plays various
important r\^oles in constraining supersymmetric models.}

\FullConference{The 15th International Conference on B-Physics at Frontier Machines at the University of Edinburgh,\\
		14 -18 July, 2014\\
		University of Edinburgh, UK}

\begin{document}

\section{Introduction}

\begin{figure}[htb]
\centering
\includegraphics[height=3in]{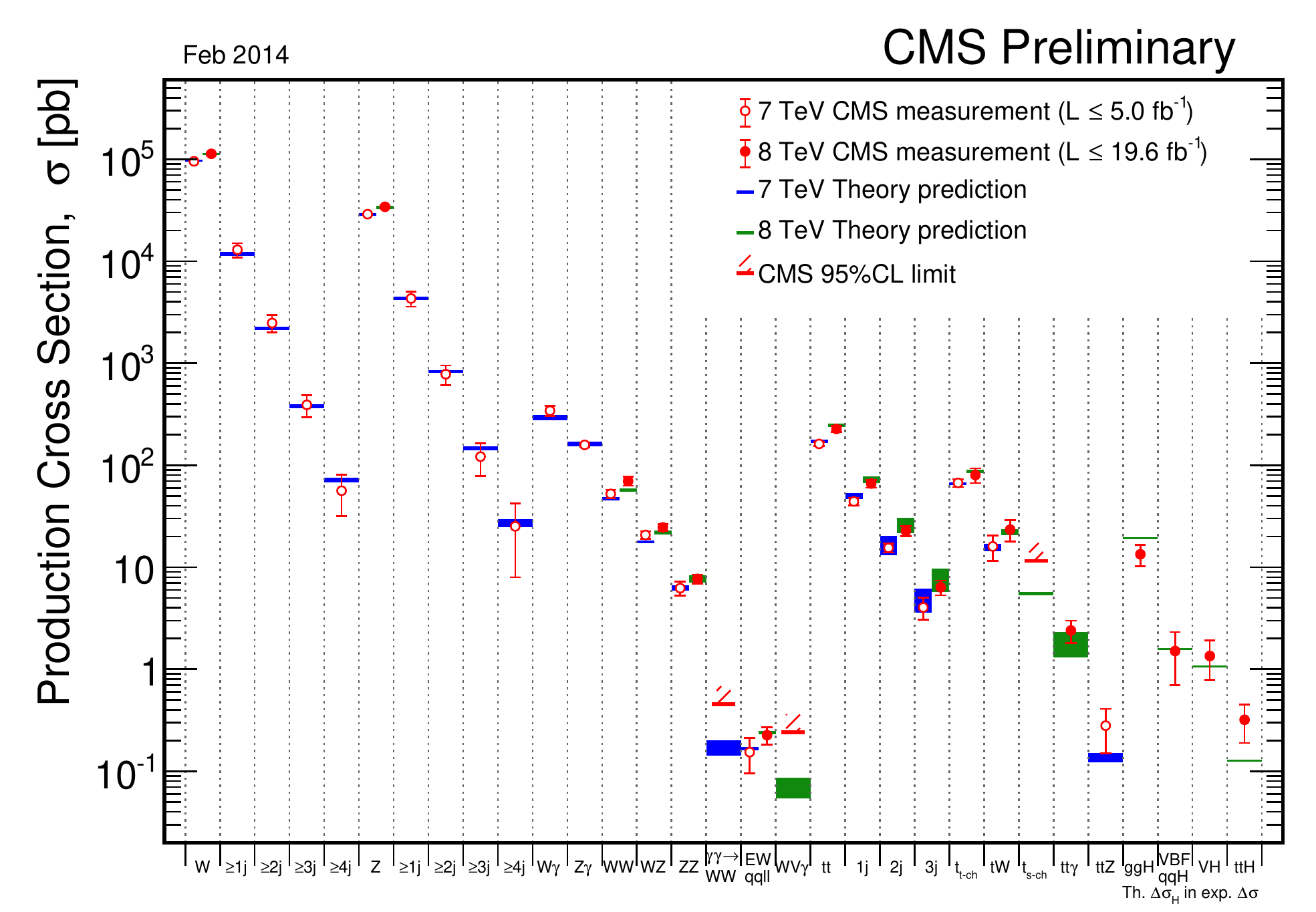}
\caption{\it A compilation of cross sections at the LHC measured by the CMS Collaboration~\protect\cite{CMSxs}.}
\label{fig:sections}
\end{figure}

The Standard Model (SM) has reigned supreme during Run 1 of the LHC. As seen in Fig.~\ref{fig:sections},
the SM has predicted successfully many cross sections for particle and jet
production at the LHC~\cite{CMSxs}. In addition to pure QCD jet production cross sections, which agree with
SM predictions over large ranges in energy and many orders of magnitude, there have been
impressive measurements of single and multiple $W^\pm$ and $Z^0$ production, as well as
multiple measurements of top quark production, both pair and single and in association with
vector bosons.

%%%%%%%%%%%%%%%%%%%%%%%%%%%%%%%%%%%%%%%%%%%%%%%%%%%%%%%%%%%%%%%%%%%%%%%%%
%%
%%   use this format to include an .eps figure into your paper
%%
%%%%%%%%%%%%%%%%%%%%%%%%%%%%%%%%%%%%%%%%%%%%%%%%%%%%%%%%%%%%%%%%%%%%%%%%%%%

Of course, the highlight of Run 1 of the LHC was the discovery by CMS and ATLAS of a (the?) Higgs boson~\cite{Higgs},
whose production has by
now been observed in three different production channels, as also seen in Fig.~\ref{fig:sections}.
The second highlight of Run 1 was the observation by LHCb and CMS of $B_s \to \mu^+ \mu^-$ decay~\cite{Bsmumu},
again in agreement with the SM.

Lovers of physics beyond the Standard Model (BSM) have had to be patient so far, though
Run~1 of the LHC has produced a few, not very significant, anomalies and excesses to 
get excited about, including in flavour physics. One of the focuses during Run~2 will be the
more detailed study of the Higgs boson and probes whether its properties deviate from
SM predictions, e.g., in the flavour sector. As I discuss later, the
measurement of the Higgs mass has produced a new reason to expect BSM physics, and the search
for BSM physics will start anew at Run~2, with its greatly increased centre-of-mass energy and
increased integrated luminosity. My personal favourite candidate for BSM physics is supersymmetry
(SUSY), and I also discuss later in this talk how SUSY models are constrained by flavour physics, as well as by the
observations to date of the Higgs boson and searches for BSM physics with Run-1 data.

\section{Higgs Physics}

The most fundamental property of the Higgs boson is its mass~\footnote{Disregarding its spin and parity,
which have by now been determined as zero and dominantly CP-even with a high degree of confidence~\cite{Hspin}, though some 
channel-dependent admixtures of CP-odd couplings are possible.}, which can be measured most accurately
in the $\gamma \gamma$ and $Z Z^* \to 2 \ell^+ 2 \ell^-$ final states. ATLAS and CMS
report accurate measurements in both these final states. ATLAS measures~\cite{ATLASmH}
\begin{eqnarray}
H \to \gamma \gamma: m_H & = \; 125.98 \pm 0.42 \pm 0.28~{\rm GeV} = \; 125.98 \pm 0.50~{\rm GeV}\, , \nonumber \\
H \to Z Z^*: m_H & = \; 125.51 \pm 0.52 \pm 0.04~{\rm GeV} = \; 125.51 \pm 0.52~{\rm GeV}\, , \nonumber \\
{\rm ATLAS~combined:} ~m_H & = \; 125.36 \pm 0.37 \pm 0.18~{\rm GeV} \; = \; 125.36 \pm 0.41~{\rm GeV}\, ,
\label{ATLASm}
\end{eqnarray}
and CMS measures~\cite{CMSmH}
\begin{eqnarray}
H \to \gamma \gamma: m_H & = \; 124.70 \pm 0.31 \pm 0.15~{\rm GeV} \; = \; 124.70^{+0.35}_{-0.34}~{\rm GeV}\, , \nonumber \\
H \to Z Z^*: m_H & = \; 125.6 \pm 0.4 \pm 0.2~{\rm GeV}\, = \; 125.6 \pm 0.4~{\rm GeV} \, , \nonumber \\
{\rm CMS~combined:} ~m_H & = \; 125.03^{+ 0.26}_{-0.27}~^{+ 0.13}_{- 0.15}~{\rm GeV} \; = \; 125.03 \pm 0.30~{\rm GeV}\, .
\label{CMSm}
\end{eqnarray}
Some interest has been generated by the differences in the masses measured in these channels,
but these have opposite signs in the two experiments:
\begin{eqnarray}
{\rm ATLAS}: \; \Delta m_H & = \; 1.47 \pm 0.67 \pm 0.18~{\rm GeV} \, , \nonumber \\
{\rm CMS}: \; \Delta m_H & = \; - 0.9 \pm 0.4 \pm 0.2 ^{+0.34}_{-0.35}~{\rm GeV} \, ,
\label{Deltam}
\end{eqnarray}
so are presumably statistical and/or systematic artefacts.
Combining naively the ATLAS and CMS measurements yields
\begin{equation}
m_H \; = \; 125.15 \pm 0.24~{\rm GeV}.
\label{mH}
\end{equation}
In addition to being a fundamental measurement in its own right, and casting light on the possible validity of various BSM models,
the precise value of $m_H$ is also important for the stability of the electroweak vacuum in the Standard Model,
as discussed later.

\begin{figure}[htb]
\centering
\includegraphics[height=2.3in]{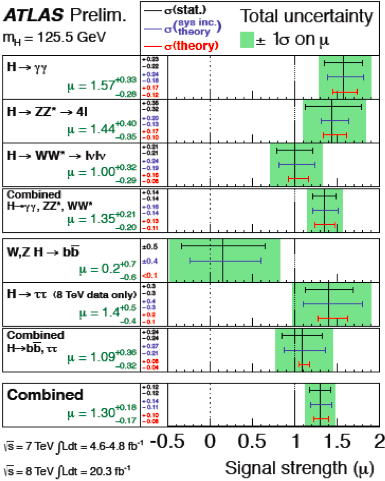}
\includegraphics[height=2.3in]{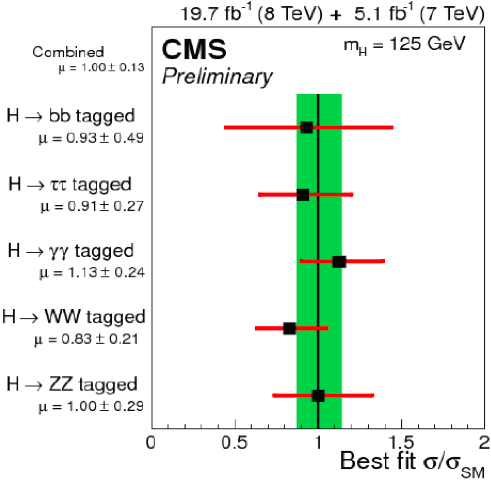}
\caption{\it The Higgs signal strengths $\mu$, normalised to unity for the SM, as measured by
ATLAS~\protect\cite{ATLASmu} (left panel) and CMS~\protect\cite{CMSmu} (right panel).}
\label{fig:strengths}
\end{figure}

As seen in Fig.~\ref{fig:strengths}, the strengths of the Higgs signals measured
by ATLAS and CMS individual channels are generally
compatible with the SM predictions within the statistical fluctuations~\cite{ATLASmu, CMSmu},
which are inevitably large at this stage. Combining their measurements
in the $\gamma \gamma$,
$Z Z^*$, $W W^*$, $b \bar{b}$ and $\tau^+ \tau^-$ channels, ATLAS
and CMS report the following overall signal strengths:
\begin{eqnarray}
{\rm ATLAS:} ~\mu & = \; 1.30 \pm 0.12 \pm 0.10 \pm 0.09 \, , \nonumber \\
{\rm CMS:} ~ \mu & = \; 1.00 \pm 0.09~^{+ 0.08}_{- 0.07} \pm 0.07 \, .
\label{mu}
\end{eqnarray}
These averages are again quite compatible with each other and with the SM,
and measurements at the Tevatron are also compatible with SM predictions for the Higgs boson~\cite{TevatronH}.

%%%%%%%%%%%%%%%%%%%%%%%%%%%%%%%%%%%%%%%%%%%%%%%%%%%%%%%%%%%%%%%%%%%%%%%%%
%%
%%   use this format to include an .eps figure into your paper
%%
%%%%%%%%%%%%%%%%%%%%%%%%%%%%%%%%%%%%%%%%%%%%%%%%%%%%%%%%%%%%%%%%%%%%%%%%%%%

\begin{figure}[htb]
\centering
\includegraphics[height=2.3in]{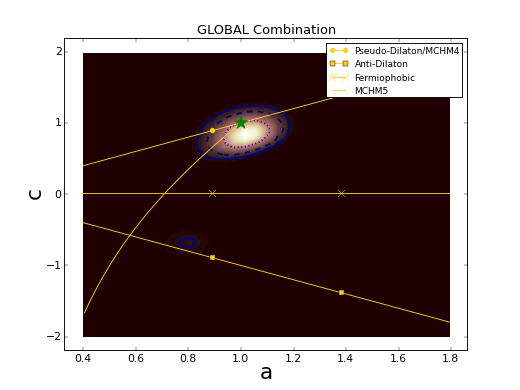}
\caption{\it A global fit to bosonic and fermionic $H$ couplings rescaled by factors $a$ and $c$, respectively,
showing the possible predictions of some composite models (yellow lines) and the SM (green star)~\protect\cite{EY3}.}
\label{fig:EY}
\end{figure}

One way to analyse the Higgs couplings is by allowing each one to differ from
the SM prediction by individual factors $\kappa_i$~\cite{HiggsxsWG}, and use the data to constrain
these factors. In the case of the Higgs couplings to fermions, this is a direct way
of probing its flavour properties. Within this general approach, it is also interesting
to impose restrictions on the $\kappa_i$ that are motivated, e.g., by specific
classes of composite models, and look for deviations from the SM that might
arise in such models. Fig.~\ref{fig:EY} shows one such example~\cite{EY3},
in which the $H$ couplings to the $W^\pm$ and $Z^0$ bosons are rescaled by a common factor $a$
and those to fermions by a common factor $c$. The data are completely consistent with the
SM case $a = c = 1$, indicated by the green star. The predictions of some specific composite
models are indicated by yellow lines: some of these models are clearly incompatible with the data,
and the survivors must be tuned to give predictions close to those of the SM.

%%%%%%%%%%%%%%%%%%%%%%%%%%%%%%%%%%%%%%%%%%%%%%%%%%%%%%%%%%%%%%%%%%%%%%%%%
%%
%%   use this format to include an .eps figure into your paper
%%
%%%%%%%%%%%%%%%%%%%%%%%%%%%%%%%%%%%%%%%%%%%%%%%%%%%%%%%%%%%%%%%%%%%%%%%%%%%

\begin{figure}[htb]
\centering
\includegraphics[height=2.15in]{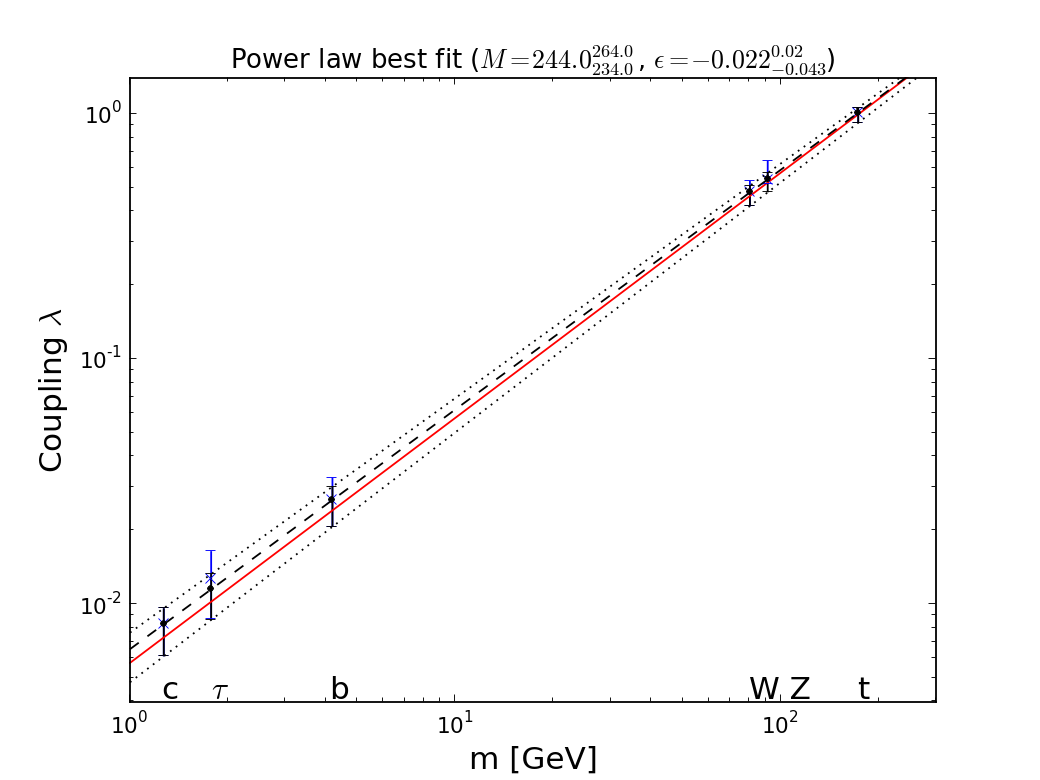}
\includegraphics[height=2.05in]{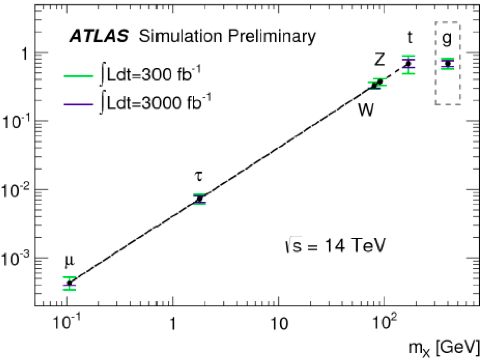}
\caption{\it Left panel: A global fit to the $H$ couplings of the form (\protect\ref{epsilon}), which is very compatible
with the expected linear mass dependence for fermions and quadratic mass dependence for bosons (red line)~\protect\cite{EY3}.
Right panel: Preliminary simulation by ATLAS of prospective fits to the couplings' mass dependence with future LHC 
data~\protect\cite{ATLASmdep}.}
\label{fig:Mass_dependence}
\end{figure}

It is a key prediction of the SM that the Higgs couplings to other particles should be related to their
masses (linearly for fermions, quadratically for bosons), and this is indirectly verified by the
measurements in Fig.~\ref{fig:strengths}. It is also possible to test this prediction directly, as seen
in Fig.~\ref{fig:Mass_dependence}. Here we made a global fit to the data then available
parametrising the Higgs couplings as~\cite{EY3}
\begin{equation}
\lambda_f \; = \; \sqrt{2} \left( \frac{m_f}{M} \right)^{(1 + \epsilon)}, \; \; 
g_V \; = \; 2 \left( \frac{M_V^{2(1 + \epsilon)}}{M^{(1 + \epsilon)}} \right) \, .
\label{epsilon}
\end{equation}
As seen in the left panel of Fig.~\ref{fig:Mass_dependence}, the data yielded
\begin{equation}
\epsilon \; = \; - 0.022^{+0.020}_{-0.043}, \; \; M \; = \; 244^{+20}_{-10}~{\rm GeV}, \,
\label{epsilonM}
\end{equation}
quite compatible with the SM predictions $\epsilon = 0$, $M = 246$~GeV.
It seems that, to a first approximation, Higgs couplings have the same flavour structure
as particle masses. The right panel shows how accurately the ATLAS Collaboration
estimates that it will be able to test the expected mass dependence of Higgs couplings
with data from future runs of the LHC~\cite{ATLASmdep}. Let us see what Run~2 will bring.

%%%%%%%%%%%%%%%%%%%%%%%%%%%%%%%%%%%%%%%%%%%%%%%%%%%%%%%%%%%%%%%%%%%%%%%%%
%%
%%   use this format to include an .eps figure into your paper
%%
%%%%%%%%%%%%%%%%%%%%%%%%%%%%%%%%%%%%%%%%%%%%%%%%%%%%%%%%%%%%%%%%%%%%%%%%%%%

\begin{figure}[h!]
\centering
\includegraphics[scale=0.5]{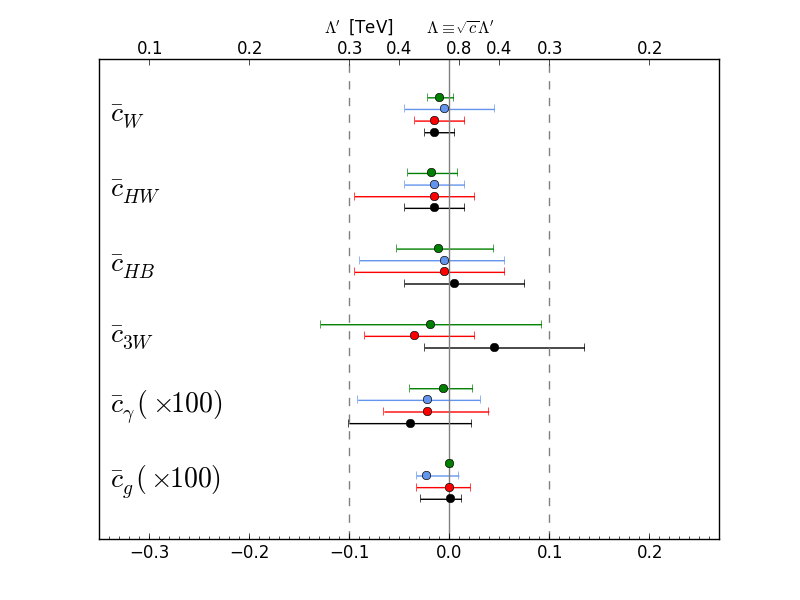}
\caption{\it The 95\% CL constraints on dimension-6 operators composed of SM fields, as
obtained from single-coefficient fits (green bars),
and the marginalised 95\% ranges for the
LHC signal-strength data combined with the kinematic distributions for associated $H + V$ production
measured by ATLAS and D0 (blue bars), combined with the LHC TGC data (red lines), and the global combination with
both the associated production and TGC data (black bars). From~\protect\cite{ESY}.}
\label{fig:ESY}
\end{figure}

Going forward, a useful way to analyse Higgs and other electroweak data in a coherent way is to use 
an `effective SM parameterisation' constructed in terms of SM fields, but including higher-dimensional
operators that might arise from integrating out heavier degrees of freedom. This opens the way to the
systematic study of electroweak precision tests and triple-gauge couplings (TGCs), as well as Higgs couplings,
in a unified and efficient framework. Some results from a recent global analyses of the LHC Run~1 constraints
on these observables is shown in Fig.~\ref{fig:ESY}~\cite{ESY}. One finds that precision electroweak measurements,
Higgs observables (including the kinematics of associated Higgs production) and TGCs play complementary
r\^oles in constraining the coefficients of possible, pushing possible new physics beyond the TeV scale in some cases.

%%%%%%%%%%%%%%%%%%%%%%%%%%%%%%%%%%%%%%%%%%%%%%%%%%%%%%%%%%%%%%%%%%%%%%%%%
%%
%%   use this format to include an .eps figure into your paper
%%

%%%%%%%%%%%%%%%%%%%%%%%%%%%%%%%%%%%%%%%%%%%%%%%%%%%%%%%%%%%%%%%%%%%%%%%%%%%

This effective field theory approach will surely be invaluable for analysing the data from LHC Run~2.

\section{Flavour Physics}

\begin{figure}[htb]
\centering
\includegraphics[height=2.3in]{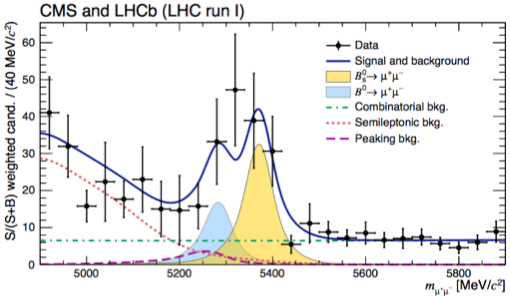}
\caption{\it The signal for $B_s \to \mu^+ \mu^-$ decay observed by the CMS and LHCb Collaborations~\protect\cite{Bsmumu},
together with a hint of $B_d \to \mu^+ \mu^-$ decay.}
\label{fig:Bmumu}
\end{figure}

The Cabibbo-Kobayashi-Maskawa (CKM) description of flavour mixing and CP violation
has made many successful predictions, and has passed most Run-1 tests with flying colours.
In particular, it predicted successfully the branching ratio for the rare decay $B_s \to \mu^+ \mu^-$~\cite{Bsmumu}:
\begin{equation}
BR(B_s \to \mu^+ \mu^-) \; = \; 2.8^{+0.7}_{-0.6} \times 10^{-9} \, ,
\label{Bsmumu}
\end{equation}
as seen in Fig.~\ref{fig:Bmumu}, which has been the second highlight of LHC Run 1. However, a point to watch during Run~2
will be the branching ratio for $B_d \to \mu^+ \mu^-$ decay, whose ratio to $B_s \to \mu^+ \mu^-$
is an ironclad prediction of the CKM model and models with minimal flavour violation (MFV),
 including many SUSY scenarios. As also seen in Fig.~\ref{fig:Bmumu},
 the joint CMS and LHCb analysis has an indication of a $B_d \to \mu^+ \mu^-$ signal
that is considerably larger than the SM prediction:
\begin{equation}
BR(B_d \to \mu^+ \mu^-) \; = \; 3.9^{+1.6}_{-1.4} \times 10^{-10} \, .
\label{Bdmumu}
\end{equation}
Could this indicate that MFV is not the whole story? {\it Une affaire \`a suivre!}

%%%%%%%%%%%%%%%%%%%%%%%%%%%%%%%%%%%%%%%%%%%%%%%%%%%%%%%%%%%%%%%%%%%%%%%%%
%%
%%   use this format to include an .eps figure into your paper
%%
%%%%%%%%%%%%%%%%%%%%%%%%%%%%%%%%%%%%%%%%%%%%%%%%%%%%%%%%%%%%%%%%%%%%%%%%%%%

\begin{figure}[htb]
\centering
\includegraphics[height=2.35in]{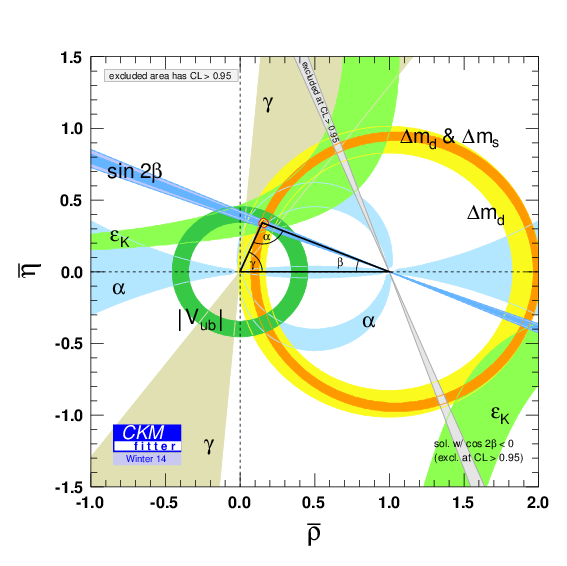}
\includegraphics[height=2.25in]{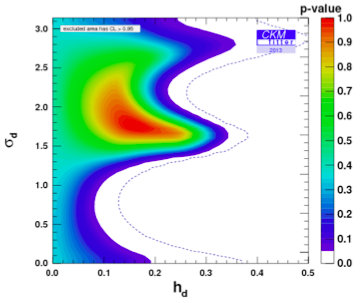}
\caption{\it Left panel: Flavour and CP violation measurements generally agree well with the CKM paradigm.
Right panel: Experimental constraint on a possible non-Standard Model contribution to $B_s$ mixing~\protect\cite{CKMFitter}.}
\label{fig:hs}
\end{figure}

Despite the successes of the CKM paradigm, see the left panel of Fig.~\ref{fig:hs},
there is scope for new physics beyond it, and some
hints of cracks in its facade. For example, the data allow an important BSM contribution to the mixing amplitude
for $B_s$ mesons: $A = A|_{SM} \times (1 + h_s e^{i \sigma_s})$,
as seen in the right panel of Fig.~\ref{fig:hs}~\cite{CKMFitter}. Also, although the early indications of CP violation in $D$ decays
above the CKM level have not been confirmed with more data, there a few intriguing anomalies.
For example, the branching ratio for $B^\pm \to \tau^\pm \nu$ decay differs from the SM prediction
by $\sim 2 \sigma$, and there are issues with $e - \mu$ universality in semileptonic $B$ decays~\cite{nonuniv}.
The most significant anomaly appears in the $P_5^\prime$ angular distribution for
$B^0 \to K^{*0} \mu^+ \mu^-$~\cite{P5prime}, though the nonperturbative corrections need to be understood
better. Also worth noting are discrepancies in the determinations of the
$V_{ub}$ CKM matrix element, and there is still an anomaly in the
diimuon asymmetry at the Tevatron~\cite{dimuon}.

%%%%%%%%%%%%%%%%%%%%%%%%%%%%%%%%%%%%%%%%%%%%%%%%%%%%%%%%%%%%%%%%%%%%%%%%%
%%
%%   use this format to include an .eps figure into your paper
%%
%%%%%%%%%%%%%%%%%%%%%%%%%%%%%%%%%%%%%%%%%%%%%%%%%%%%%%%%%%%%%%%%%%%%%%%%%%%

Some anomalies do seem to be going away, such as the forward-backward asymmetry in
$t \bar{t}$ production, which now agrees with higher-order QCD calculations~\cite{ttbarAFB}, as does the
 $t \bar{t}$ rapidity asymmetry measured at the LHC. However,
there are plenty of flavour issues to be addressed during LHC Run 2.

\begin{figure}[htb]
\centering
\includegraphics[height=2.3in]{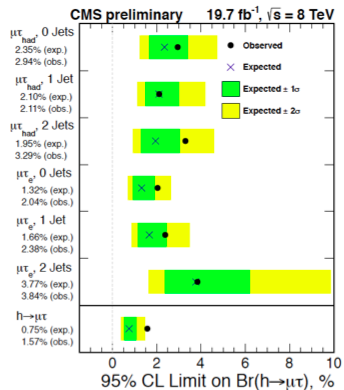}
\caption{\it Results from the CMS search for $H \to \mu \tau$ decay~\protect\cite{CMSHmutau}.}
\label{fig:Hmutau}
\end{figure}

One of the key predictions of the SM, which also holds in many SUSY models,
is that the Higgs couplings to fermions should conserve flavour to a very good
approximation, and this is consistent with the upper limits on low-energy effective
flavour-changing interactions. However, these would allow lepton-flavour-violating
Higgs couplings to fermions far above the SM predictions, so looking for such
interactions is a possible window on BSM physics. Specifically, we found that
the branching ratios for $H \to \mu \tau$ and $H \to e \tau$ decays could each be
as large as ${\cal O}(10)$\%, whereas the branching ratio for $H \to \mu e$
must be $\lesssim 10^{-5}$~\cite{BEI}. The CMS Collaboration has recently searched for
$H \to \mu \tau$ decays in various $H$ production modes, as seen in Fig.~\ref{fig:Hmutau}, and found~\cite{CMSHmutau}
\begin{equation}
{\rm BR}(H \to \mu \tau) \; = \; 0.89^{+0.40}_{-0.37} \, \% \, ,
\label{Hmutau}
\end{equation}
which has a background-only $p$-value of 0.007, corresponding to 2.46$\sigma$.
On the one hand, this tells us that Higgs measurements are already probing
flavour physics beyond the previous low-energy experiments and, on the other
hand, it will be very interesting to see corresponding results from ATLAS and results from Run~2!

%%%%%%%%%%%%%%%%%%%%%%%%%%%%%%%%%%%%%%%%%%%%%%%%%%%%%%%%%%%%%%%%%%%%%%%%%
%%
%%   use this format to include an .eps figure into your paper
%%
%%%%%%%%%%%%%%%%%%%%%%%%%%%%%%%%%%%%%%%%%%%%%%%%%%%%%%%%%%%%%%%%%%%%%%%%%%%

\begin{figure}[htb]
\centering
\includegraphics[height=2.3in]{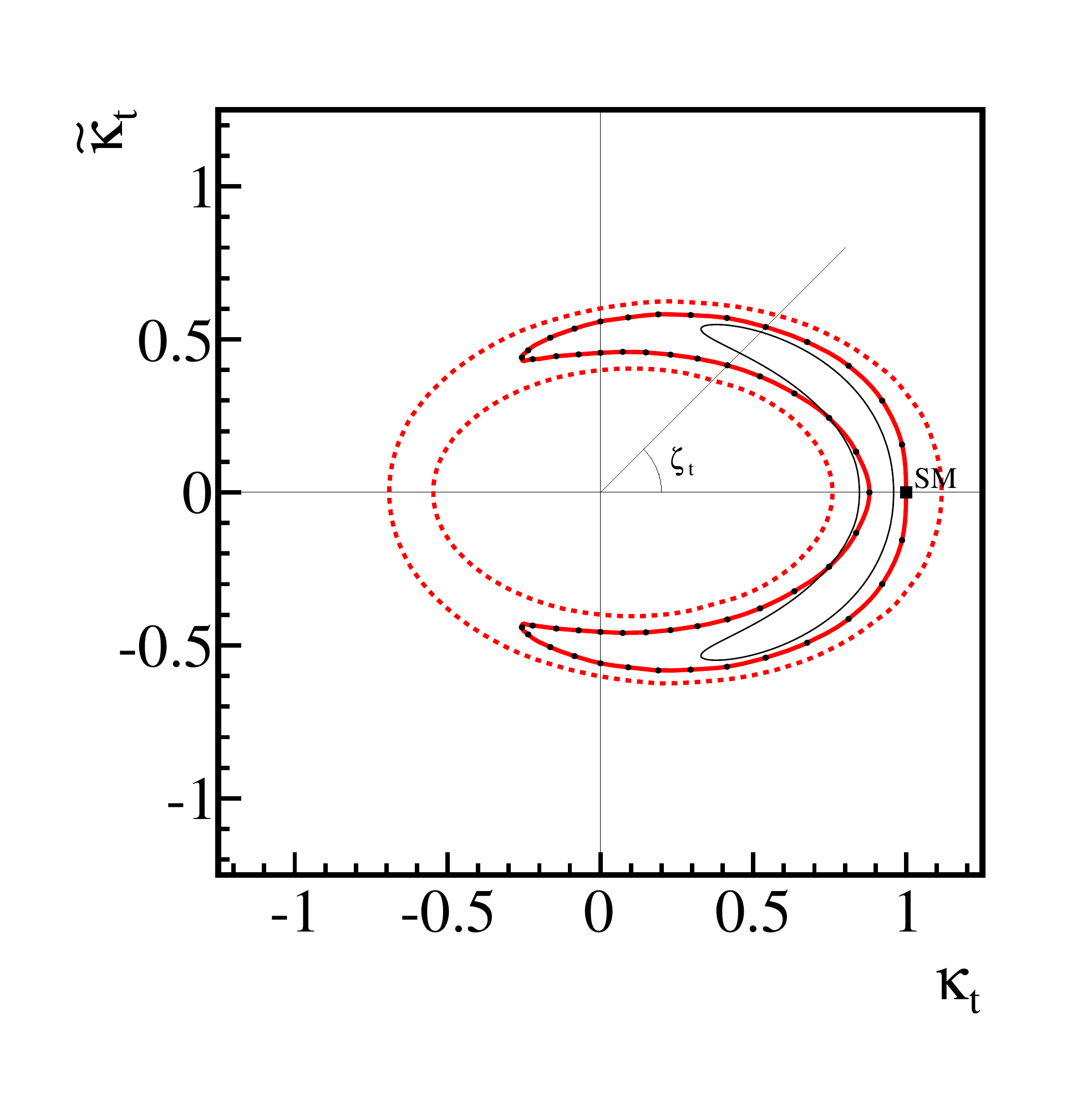}
\includegraphics[height=2.6in]{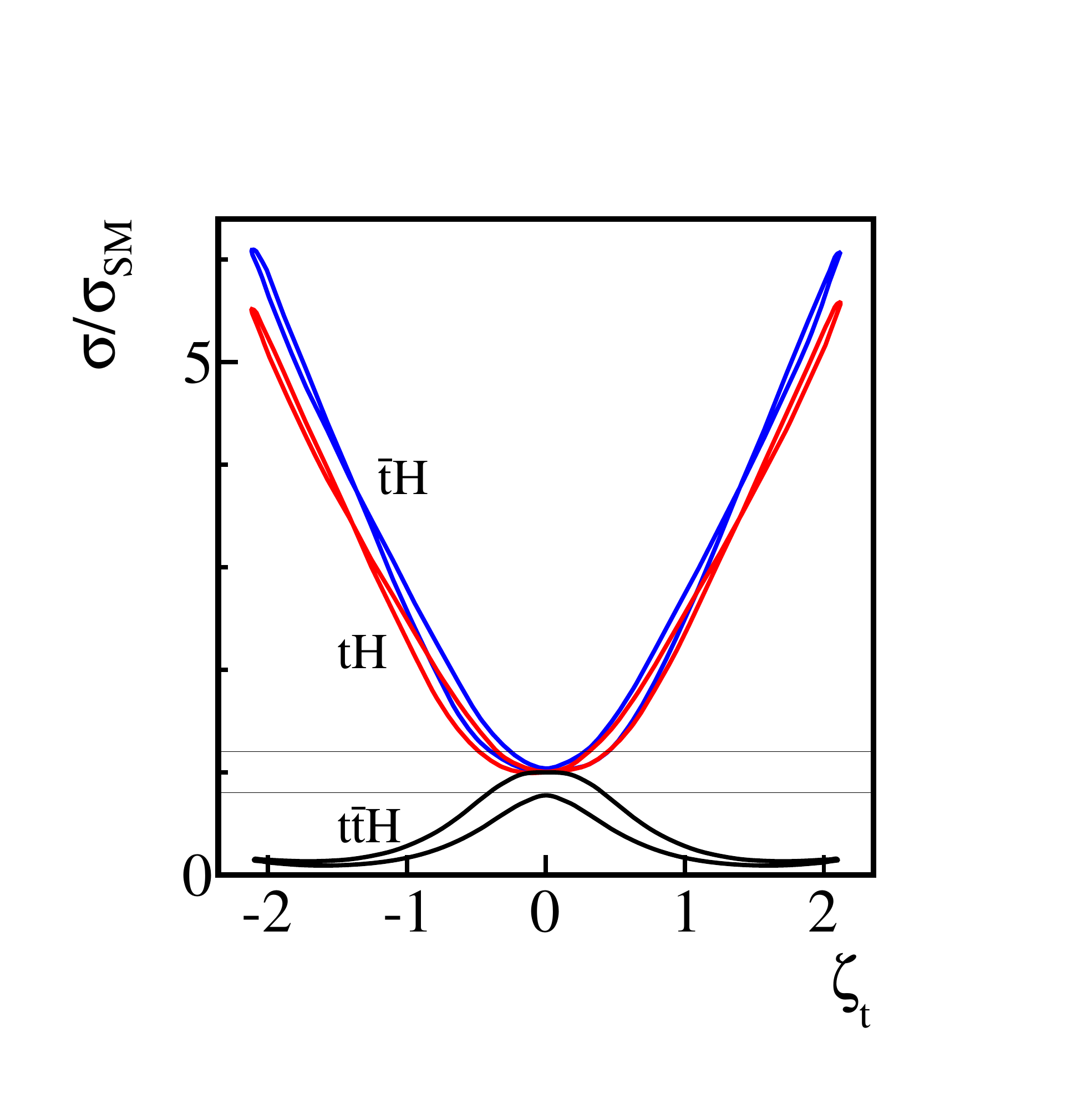}
\caption{\it Left panel: The constraints on the CP-even and -odd $H t {\bar t}$ couplings from
measurements of the $H g g$ and $H \gamma \gamma$ couplings. Right panel: The dependences of
the cross sections for associated
${\bar t} t H$, $t H$ and ${\bar t} H$ production, on $\zeta_t \equiv \arg({\rm CP-odd~coupling}/{\rm CP-even~coupling})$.
From~\protect\cite{EHST}.}
\label{fig:EHST}
\end{figure}

It will also be interesting to see whether the Higgs couplings contain CP-odd
admixtures. We already know that its couplings to massive gauge bosons are
predominantly CP-even, but we have very little direct information about its
couplings to fermions, though there have been suggestions how to measure
the CP properties of the $H \tau^+ \tau^-$ coupling, for example. How about the
$H t {\bar t}$ coupling? There are indirect constraints from the available experimental
information on the $H g g$ and $H \gamma \gamma$ couplings, as seen in the left panel of Fig.~\ref{fig:EHST}.
Direct information could come in the future from measurements of the cross sections for associated
${\bar t} t H$, $t H$ and ${\bar t} H$ production, which are sensitive to $\zeta_t \equiv \arg({\rm CP-odd~coupling}/{\rm CP-even~coupling})$, as seen in Fig.~\ref{fig:EHST}~\cite{EHST}. One could also look for
CP-violating final-state asymmetries in $t ({\bar t})$ decays. A challenge for Run~2 and beyond!

%%%%%%%%%%%%%%%%%%%%%%%%%%%%%%%%%%%%%%%%%%%%%%%%%%%%%%%%%%%%%%%%%%%%%%%%%
%%
%%   use this format to include an .eps figure into your paper
%%
%%%%%%%%%%%%%%%%%%%%%%%%%%%%%%%%%%%%%%%%%%%%%%%%%%%%%%%%%%%%%%%%%%%%%%%%%%%

\section{The SM is not enough!}

Now that the SM has apparently been completed by the discovery of a (the?)
Higgs boson, some might argue that there is no physics beyond the SM.
However, history is lettered with distinguished physicists (and others) who declared
 {\it ``game over"} prematurely. Albert Michelson declared in 1894 that {\it ``The more important
fundamental laws and facts of physical science have all been discovered"}, just 
before the discoveries of radioactivity and the electron. Lord Kelvin declared in 1900 that
{\it ``There is nothing new to be discovered in physics now,
all that remains is more and more precise measurement"}, just before Einstein
postulated the photon and proposed special relativity.

There are many reasons why the SM is not enough. Inspired by James Bond~\cite{Bond},
here I just mention 007 of them. 1) Taking at face value
the measured values of $m_t$ and $m_H$, the electroweak vacuum is {\it probably} unstable, unless additional physics intervenes.
2) The dark matter required by astrophysics and cosmology cannot be provided by the SM.
3) The origin of matter in the Universe requires addition CP violation beyond the CKM model.
4) Explaining the small sizes of the neutrino masses requires physics beyond the SM.
5) The hierarchy and fine-tuning problems suggest there is new physics at the TeV scale.
6) Cosmological inflation (probably) requires physics beyond the SM, in particular 
because the effective Higgs potential is probably negative at high scales, as discussed shortly.
7) The construction of a consistent quantum theory of gravity certainly involves going (far) beyond the SM.

\section{New Reasons to Love SUSY}

Most of these issues would be at least alleviated by supersymmetry, and the LHC Run~1
has given us at least three new reasons to love SUSY, as I now discuss.

\subsection{The Instability of the Electroweak Vacuum}

In the SM the effective electroweak potential resembles a Mexican hat,
invariant under the SM SU(2)$\times$U(1) symmetry, but unstable at the origin. The electroweak vacuum lies in a surrounding valley 
where $\langle H \rangle \equiv v = 246$~GeV. Beyond this valley, the brim of the hat rises, but how far?
Calculations show that renormalization by the top quark overwhelms that by the Higgs itself,
for the measured values of $m_t$ and $m_H$, turning the brim 
down at large field values. Consequently, the present electroweak vacuum is in principle
unstable, potentially collapsing into an anti-de-Sitter 'Big Crunch' via quantum tunnelling though the brim.

\begin{figure}[htb]
\centering
\includegraphics[height=2.1in]{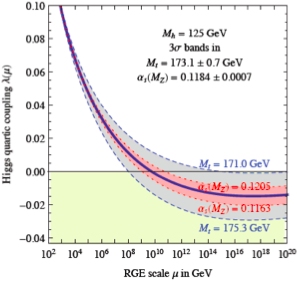}
\includegraphics[height=2.1in]{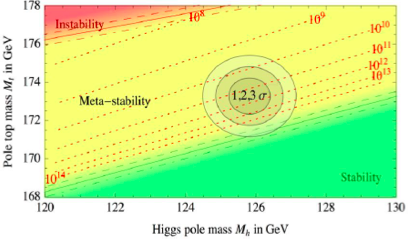}
\caption{\it Left panel: Within the SM, normalisation by the top quark appears to drive the Higgs self-coupling $\lambda < 0$.
Right panel: The regions of vacuum stability, metastability and instability in the $(m_H, m_t)$ plane.
Both panels are from~\protect\cite{Buttazzo}.}
\label{fig:Buttazzo}
\end{figure}

According to the best SM calculations available, shown in the left panel of Fig.~\ref{fig:Buttazzo},
the brim turns down at a Higgs scale $\Lambda$~\cite{Buttazzo}:
\begin{equation}
\log_{10} \left( \frac{\Lambda}{{\rm GeV}} \right) \; = \; 11.3 + 1.0 \left(\frac{m_H}{{\rm GeV}} - 125.66 \right)
- 1.2 \left( \frac{m_t}{{\rm GeV}} - 173.10 \right) + 0.4 \left(\frac{\alpha_s(M_Z) - 0.1184}{0.0007} \right) \, .
\label{Buttazzo}
\end{equation}
Using the official world average values of $m_t$, $m_H$ and $\alpha_s (M_Z)$, one may estimate
\begin{equation}
\Lambda \; = \; 10^{10.5 \pm 1.1}~{\rm GeV}
\label{Lambda}
\end{equation}
though the error is neither Gaussian nor symmetric. As seen in the right panel of Fig.~\ref{fig:Buttazzo}, this
calculation is most sensitive to $m_t$. Subsequent to the determination of the world average,
D0 has reported a new, higher, value of $m_t$~\cite{D0mt}, which would tend to decrease
$\log_{10} (\Lambda/{\rm GeV})$ by 2.0, further destabilising the electroweak vacuum, but CMS has
reported new analyses of $m_t$~\cite{CMSmt} that would {\it increase} $\log_{10} (\Lambda/{\rm GeV})$ by 1.6, making
the vacuum more stable. A more accurate value of $m_t$ would fix the fate of the Universe within the SM.
However, the experimental effort must be matched by better theoretical understanding of the relationship
between the effective mass parameter used by experiments in their Monte Carlos and the
parameter $m_t$ in the SM Lagrangian~\cite{Moch}.

%%%%%%%%%%%%%%%%%%%%%%%%%%%%%%%%%%%%%%%%%%%%%%%%%%%%%%%%%%%%%%%%%%%%%%%%%
%%
%%   use this format to include an .eps figure into your paper
%%
%%%%%%%%%%%%%%%%%%%%%%%%%%%%%%%%%%%%%%%%%%%%%%%%%%%%%%%%%%%%%%%%%%%%%%%%%%%

In much of the favoured parameter space, the electroweak vacuum
would (probably) live longer than the age of the Universe, so you might be tempted to shrug your
shoulders, but there is another problem.
Observations of the cosmic microwave background suggest that the Universe once had a very
higher energy density during an inflationary epoch~\cite{CMB}. Quantum and thermal
fluctuations during this epoch would have favoured a transition away from the electroweak minimum and towards
an anti-de-Sitter `Big Crunch' region~\cite{Oops}. One could argue that
a non-anti-de-Sitter region containing us might have survived. Alternatively, the
problem could be avoided in the presence of higher-dimensional
terms in the effective potential~\cite{Sher}. This is just one example of new physics beyond the SM
that could have averted this cosmological disaster: another is supersymmetry, which would have
prevented the brim of the Mexican hat from turning down in the first place.

\subsection{The Higgs Mass}

It is well known that SUSY predicted correctly that the mass of the Higgs boson
should be $\lesssim 130$~GeV in simple models~\cite{SUSYmH}. This is because SUSY
specifies the quartic Higgs coupling before renormalisation. Loop corrections
to $m_H$ due (in particular) to $m_t$ increase it significantly, but are under control and accurate to with an estimated 
uncertainty of $\pm 3$~GeV for given values of the SUSY input parameters~\cite{FH}.

\subsection{Higgs Couplings}

Since there are two physical neutral CP-even Higgs bosons in the minimal SUSY
extension of the SM (MSSM), and a neutral CP-even boson $A$ that could mix
with them in the presence of CP violation, the couplings of the discovered Higgs
boson could, in principle, have differed significantly from those of the SM Higgs
boson. However, since around 2001 it has been known~\cite{EHOW} that constraints from
LEP and $b \to s \gamma$ already implied within simple SUSY models that the
Higgs couplings would be very similar to those in the SM, as observed.

\begin{figure}[htb]
\centering
\includegraphics[height=3in]{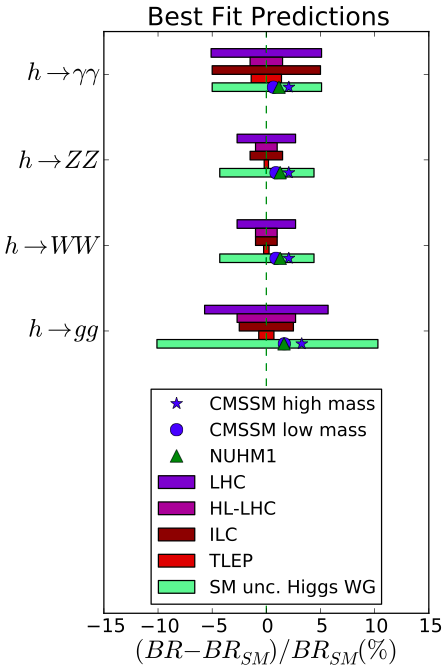}
\caption{\it Comparison of the present precisions in measurements of various Higgs branching ratios,
prospective measurements at future colliders, the current theoretical uncertainties, and the (small)
deviations from the SM predictions found at the best-fit points in various SUSY models. From~\protect\cite{TLEP}.}
\label{fig:TLEP}
\end{figure}

In the cases of contemporary best fits to the LHC and other data within simple SUSY models with all SUSY-breaking
parameters constrained to be equal at the GUT scale (the CMSSM), or allowing
one or two degrees of non-universality in the soft SUSY-breaking contributions
to Higgs masses (NUHM1 and NUHM2), one finds that Higgs couplings should be
much closer to the SM predictions than the experimental and theoretical uncertainties would allow.
As seen in Fig.~\ref{fig:TLEP}, it would require a high-luminosity circular $e^+ e^-$ collider
to distinguish these model predictions from the SM~\cite{TLEP}.

%%%%%%%%%%%%%%%%%%%%%%%%%%%%%%%%%%%%%%%%%%%%%%%%%%%%%%%%%%%%%%%%%%%%%%%%%
%%
%%   use this format to include an .eps figure into your paper
%%
%%%%%%%%%%%%%%%%%%%%%%%%%%%%%%%%%%%%%%%%%%%%%%%%%%%%%%%%%%%%%%%%%%%%%%%%%%%

\subsection{Not forgetting ...}

... the many reasons for loving SUSY established earlier, such as 
alleviating the fine-tuning aspect of the hierarchy problem, providing
a natural candidate for the cold dark matter, facilitating grand unification
and playing an essential r\^ole in string theory. It would be a shame if Nature
did not succumb to SUSY's manifold charms.

\section{Supersymmetry}

\begin{figure}[htb]
\centering
\includegraphics[height=2.3in]{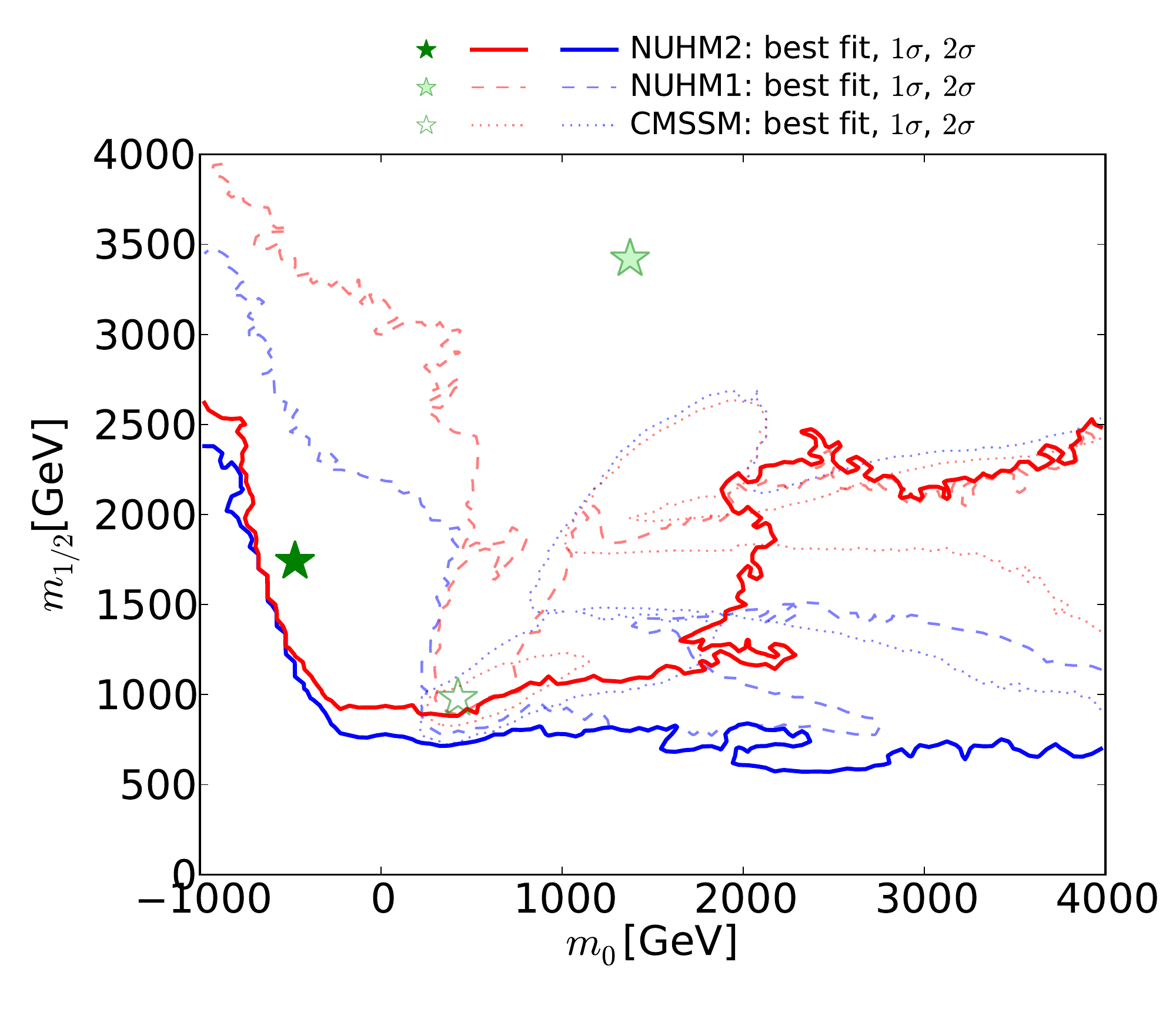}
\includegraphics[height=2.2in]{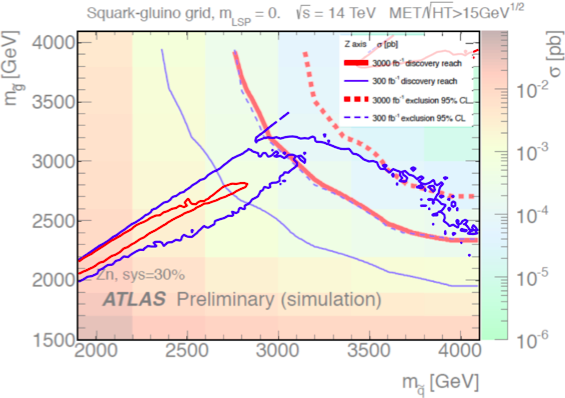}
\caption{\it Left panel: The 68\% CL (red) and 95\% CL contours (blue) in the $(m_0, m_{1/2})$ panes
 for the CMSSM (dotted lines), NUHM1 (dashed lines) and NUHM2 (solid lines)~\protect\cite{MC11}.
 Right panel: The reach of ATLAS in the $(m_{\tilde q}, m_{\tilde g})$ plane for exclusion and discovery
 with 300 and 3000/fb of integrated LHC liminosity at high energy~\protect\cite{ATLASmdep},
 compared with the 68\% CL (red) and 95\% regions in the CMSSM.}
\label{fig:MC10}
\end{figure}

Despite our ardent love for SUSY, so far she has been very coy. Direct searches for SUSY at the LHC have
drawn blanks so far. This is also the case for searches for the scattering of dark matter particles,
indirect searches in flavour physics, etc.. The results of global fits to the CMSSM, NUHM1 and NUHM2,
combining these constraints and requiring that the relic supersymmetric particle density be
within the cosmological range, are shown projected on the $(m_0, m_{1/2})$ plane
in the left panel of Fig.~\ref{fig:MC10}~\cite{MC10, MC11}. The right panel of Fig.~\ref{fig:MC10} displays the
$(m_{\tilde q}, m_{\tilde g})$ plane, showing prospective exclusion and discovery reaches of the LHC
in future runs with 300 and 3000/fb of luminosity at high energy~\cite{ATLASmdep}. Superposed on this plane are the 68
and 95\% CL contours found in the global fit to the CMSSM. As already seen in the left panel of
Fig.~\ref{fig:MC10}, there are two distinct regions, the lower-mass one being favoured by the
disagreement between experiment and the SM prediction of $g_\mu - 2$.
We see that the LHC could detect squarks and gluinos if Nature is described by
supersymmetry with parameters in this lower-mass region. 

%%%%%%%%%%%%%%%%%%%%%%%%%%%%%%%%%%%%%%%%%%%%%%%%%%%%%%%%%%%%%%%%%%%%%%%%%
%%
%%   use this format to include an .eps figure into your paper
%%
%%%%%%%%%%%%%%%%%%%%%%%%%%%%%%%%%%%%%%%%%%%%%%%%%%%%%%%%%%%%%%%%%%%%%%%%%%%

\begin{figure}[htb]
\centering
\includegraphics[height=2.1in]{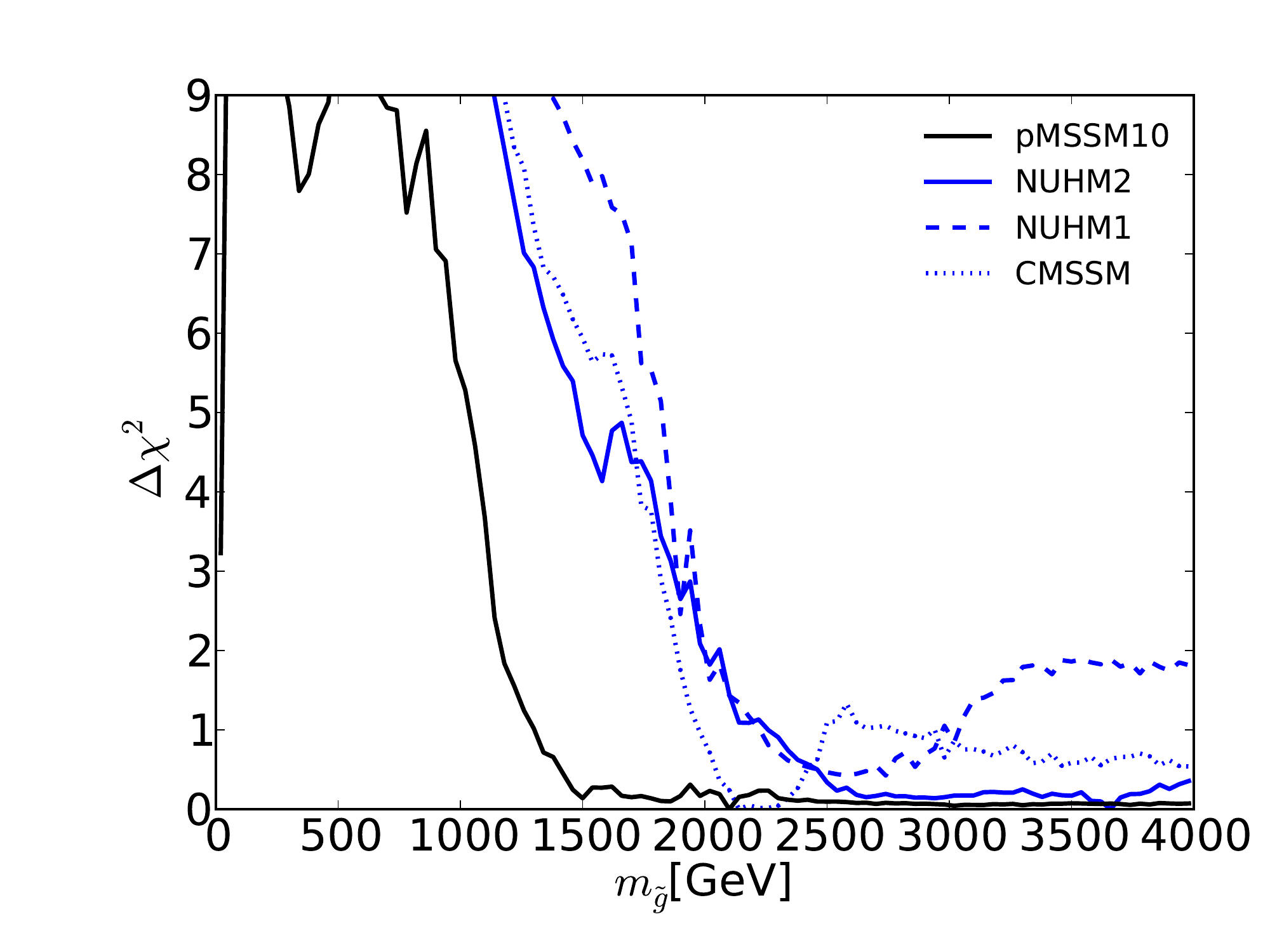}
\includegraphics[height=2.1in]{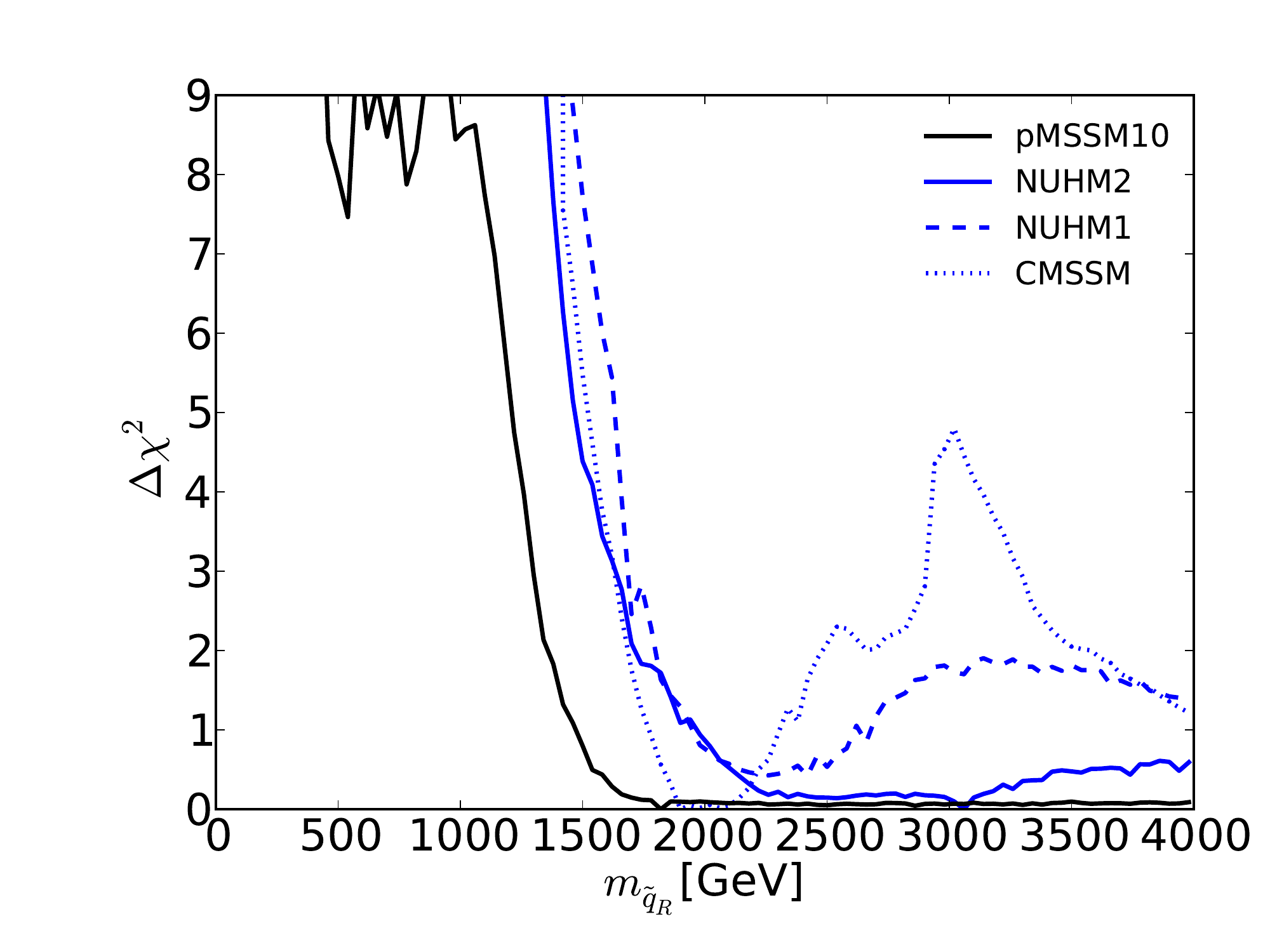}
\caption{\it The one-dimensional $\chi^2$ likelihood function for the mass of the gluino (left panel) and the right-handed squark
(right panel) in the CMSSM, NUHM1, NUHM2 and pMSSM~\protect\cite{MC11}.}
\label{fig:MC11}
\end{figure}

Run 1 of the LHC imposed strong limits on strongly-interacting sparticles, whereas the limits
on electroweakly-interacting sparticles are significantly weaker. It is only in models with universality
imposed on the soft SUSY-breaking contributions to sparticle masses that the squark and gluino limits
lead to strong limits on the masses of electroweakly-interacting sparticles. One may, instead,
consider the phenomenological MSSM (pMSSM) in which no universality is assumed. In this case,
the the lower limits on the gluino and squark masses are reduced, compared with the CMSSM, NUHM1
and NUHM2, as seen in Fig.~\ref{fig:MC11}, enhancing the prospects for discovering SUSY in LHC Run~2~\cite{MC11}.

%%%%%%%%%%%%%%%%%%%%%%%%%%%%%%%%%%%%%%%%%%%%%%%%%%%%%%%%%%%%%%%%%%%%%%%%%
%%
%%   use this format to include an .eps figure into your paper
%%
%%%%%%%%%%%%%%%%%%%%%%%%%%%%%%%%%%%%%%%%%%%%%%%%%%%%%%%%%%%%%%%%%%%%%%%%%%%

\begin{figure}[htb]
\centering
\includegraphics[height=2.0in]{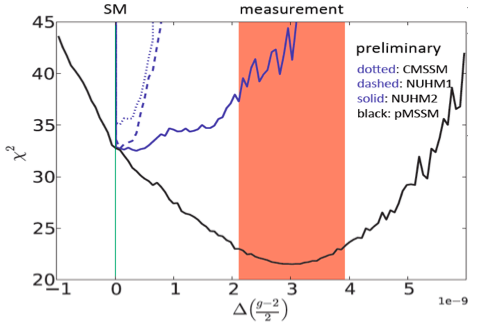}
\includegraphics[height=2.1in]{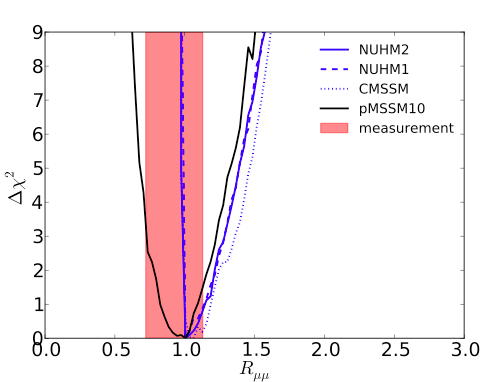}
\caption{\it The one-dimensional $\chi^2$ likelihood function for $g_\mu - 2$ (left panel) and $B_s \to \mu^+ \mu^-$
(right panel) in the CMSSM, NUHM1, NUHM2 and pMSSM~\protect\cite{MC11}.}
\label{fig:g-2}
\end{figure}

One intriguing feature of the pMSSM is that the decoupling of the masses of the strongly- and electroweakly-interacting
sparticles revives the possibility that supersymmetry could explain the discrepancy between the
experimental measurement of $g_\mu - 2$ and the value calculated in the SM. As seen in the left panel of Fig.~\ref{fig:g-2},
the LHC constraints imply that the CMSSM, NUHM1 and NUHM2 all predict values of the $g_\mu - 2$
that are very similar to the SM prediction, whereas the pMSSM could accommodate the experimental
measurement. It is good that two new experiments to measure $g_\mu - 2$ are being planned~\cite{futureg-2},
and that other experiments will enable the SM predictions to be refined and hence the discrepancy between
the SM and experiment to be clarified.

%%%%%%%%%%%%%%%%%%%%%%%%%%%%%%%%%%%%%%%%%%%%%%%%%%%%%%%%%%%%%%%%%%%%%%%%%
%%
%%   use this format to include an .eps figure into your paper
%%
%%%%%%%%%%%%%%%%%%%%%%%%%%%%%%%%%%%%%%%%%%%%%%%%%%%%%%%%%%%%%%%%%%%%%%%%%%%

On the other hand, as seen in the right panel of Fig.~\ref{fig:g-2}, all of the CMSSM, NUHM1, NUHM2 and pMSSM
can accommodate the experimental measurement of $B_s \to \mu^+ \mu^-$. Interestingly, the pMSSM is
the only one of these models that could yield a value of the branching ratio for this decay that is
any lower than in the SM.

There have been spikes of interest in a couple of hints in the Run~1 data of signals that might
be due to the production of electroweak sparticles. One was an apparent enhancement of the
$W^+ W^-$ cross section above the SM prediction~\cite{highWW}, and the other was a possible `edge' effect
in the $\mu^+ \mu^-$ invariant-mass distribution found by CMS~\cite{CMSedge}. However, the significance of the $W^+ W^-$
cross-section discrepancy is much reduced when NNLO QCD effects are taken into account~\cite{WWNNLO}, and the
dilepton edge effect should be given the opportunity to grow when more data are accumulated.
Let us wait to see what Run~2 of the LHC brings.

\section{Theoretical Perplexity}

On the one hand, the discovery of a (the?) Higgs boson during Run 1 of the LHC has
been a subject for rejoicing. On the other hand, the absence of any hard direct
evidence for new physics beyond the SM, coupled with the apparently (bog) standard
nature of the Higgs, has left theorists perplexed which BSM horse to back. (For the
experimentalists, it is clear: more energy and more luminosity!)

Despite the lack of any evidence for SUSY, I would say that it has
fared less badly than some rival theories. For example, generic composite models
could have accommodated differences fem SM Higgs properties that are larger than
those allowed by Run~1 data, whereas SUSY predicted successfully both its
mass and the SM-ness of its couplings. Siren voices may sing the praises of split
SUSY, high-scale SUSY, modifying or abandoning naturalness, or embracing the
string landscape. My point of view, however, while admitting the need for new ideas,
is that SUSY anywhere is better than SUSY nowhere. Let us see what Run~2 brings!

Perhaps the most significant hint from LHC Run~1 has been the fact that
$(m_H, m_t)$ apparently lie within the zone where extrapolation to high
scales indicates that the electroweak vacuum is unstable - though close
to the boundary between the zones of stability and instability. If these
parameters do indeed lie within the unstable region, this would provide
a strong argument for new physics, as discussed earlier. On the other hand, if these masses
do lie on or close to the stability boundary, perhaps there is some critical
phenomenon to be understood? We trust that Run~2 will clarify where
we are located in the $(m_H, m_t)$ plane.

\section{Patience!}

After its proposal, it took 48 years for the Higgs boson to be discovered, a time-lag
that was longer than those between the proposals and discoveries of other
elementary particles~\cite{Economist}. So lovers of SUSY can be patient. At the time of writing, only
41 years have passed since the first proposal of four-dimensional
supersymmetric field theories. If SUSY is discovered during LHC Run~2 or 3,
the SUSY time-lag will be less than that for the Higgs boson. Let us see
what happens when the LHC restarts!\\

%%  if necessary
%\Acknowledgements
%Work supported by the European Research Council 
%via the Advanced Investigator Grant 267352 and by STFC (UK) via the research grant ST/J002798/1.


\begin{thebibliography}{99}

\bibitem{CMSxs}
CMS Collaboration, {\tt https://twiki.cern.ch/twiki/bin/view/CMSPublic/PhysicsResultsSMP}.

\bibitem{Higgs}
G.~Aad {\it et al.}  [ATLAS Collaboration],
  %``Observation of a new particle in the search for the Standard Model Higgs boson with the ATLAS detector at the LHC,''
  Phys.\ Lett.\ B {\bf 716} (2012) 1
  [arXiv:1207.7214 [hep-ex]];
  %%CITATION = ARXIV:1207.7214;%%
  %3633 citations counted in INSPIRE as of 08 Dec 2014
  S.~Chatrchyan {\it et al.}  [CMS Collaboration],
  %``Observation of a new boson at a mass of 125 GeV with the CMS experiment at the LHC,''
  Phys.\ Lett.\ B {\bf 716} (2012) 30
  [arXiv:1207.7235 [hep-ex]].
  %%CITATION = ARXIV:1207.7235;%%
  %3567 citations counted in INSPIRE as of 08 Dec 2014
  
\bibitem{Bsmumu}
V.~Khachatryan {\it et al.}  [CMS and LHCb Collaborations],
  %``Observation of the rare $B^0_s\to\mu^+\mu^-$ decay from the combined analysis of CMS and LHCb data,''
  arXiv:1411.4413 [hep-ex] and references therein.
  %%CITATION = ARXIV:1411.4413;%%
  %4 citations counted in INSPIRE as of 08 Dec 2014
  
\bibitem{Hspin}
G.~Aad {\it et al.}  [ATLAS Collaboration],
  %``Evidence for the spin-0 nature of the Higgs boson using ATLAS data,''
  Phys.\ Lett.\ B {\bf 726} (2013) 120
  [arXiv:1307.1432 [hep-ex]];
  %%CITATION = ARXIV:1307.1432;%%
  %199 citations counted in INSPIRE as of 08 Dec 2014
  V.~Khachatryan {\it et al.}  [CMS Collaboration],
  %``Constraints on the spin-parity and anomalous HVV couplings of the Higgs boson in proton collisions at 7 and 8 TeV,''
  arXiv:1411.3441 [hep-ex].
  %%CITATION = ARXIV:1411.3441;%%
  
\bibitem{ATLASmH}
G.~Aad {\it et al.}  [ATLAS Collaboration],
  %``Measurement of the Higgs boson mass from the $H\rightarrow \gamma\gamma$ and $H \rightarrow ZZ^{*} \rightarrow 4\ell$ channels with the ATLAS detector using 25 fb$^{-1}$ of $pp$ collision data,''
  Phys.\ Rev.\ D {\bf 90} (2014) 052004
  [arXiv:1406.3827 [hep-ex]].
  %%CITATION = ARXIV:1406.3827;%%
  %74 citations counted in INSPIRE as of 08 Dec 2014
  
\bibitem{CMSmH}
CMS Collaboration, {\tt http://cds.cern.ch/record/1728249/files/HIG-14-009-pas.pdf}.

\bibitem{ATLASmu}
ATLAS Collaboration, {\tt https://twiki.cern.ch/twiki/bin/view/AtlasPublic/HiggsPublicResults}.

\bibitem{CMSmu}
S.~Chatrchyan {\it et al.}  [CMS Collaboration],
  %``Observation of a new boson with mass near 125 GeV in pp collisions at $\sqrt{s}$ = 7 and 8 TeV,''
  JHEP {\bf 1306} (2013) 081
  [arXiv:1303.4571 [hep-ex]].
  %%CITATION = ARXIV:1303.4571;%%
  %239 citations counted in INSPIRE as of 08 Dec 2014
  
\bibitem{TevatronH}
CDF and D0 Collaborations, {\tt http://tevnphwg.fnal.gov}.

\bibitem{HiggsxsWG}
A.~David {\it et al.}  [LHC Higgs Cross Section Working Group Collaboration],
  %``LHC HXSWG interim recommendations to explore the coupling structure of a Higgs-like particle,''
  arXiv:1209.0040 [hep-ph].
  %%CITATION = ARXIV:1209.0040;%%
  %108 citations counted in INSPIRE as of 08 Dec 2014
  
\bibitem{EY3}
J.~Ellis and T.~You,
  %``Updated Global Analysis of Higgs Couplings,''
  JHEP {\bf 1306} (2013) 103
  [arXiv:1303.3879 [hep-ph]].
  %%CITATION = ARXIV:1303.3879;%%
  %119 citations counted in INSPIRE as of 23 Aug 2014
  
 \bibitem{ATLASmdep}
ATLAS Collaboration,
  %``Physics at a High-Luminosity LHC with ATLAS,''
  arXiv:1307.7292 [hep-ex].
  %%CITATION = ARXIV:1307.7292;%%
  %55 citations counted in INSPIRE as of 08 Dec 2014

\bibitem{ESY}
J.~Ellis, V.~Sanz and T.~You,
  %``The Effective Standard Model after LHC Run I,''
  arXiv:1410.7703 [hep-ph] and references therein.
  %%CITATION = ARXIV:1410.7703;%%
  %3 citations counted in INSPIRE as of 08 Dec 2014
      
\bibitem{CKMFitter}
J.~Charles, S.~Descotes-Genon, Z.~Ligeti, S.~Monteil, M.~Papucci and K.~Trabelsi,
  %``Future sensitivity to new physics in B_d, B_s and K mixings,''
  Phys.\ Rev.\ D {\bf 89} (2014) 033016
  [arXiv:1309.2293 [hep-ph]].
  %%CITATION = ARXIV:1309.2293;%%
  %12 citations counted in INSPIRE as of 23 Aug 2014

\bibitem{nonuniv}
R.~Aaij {\it et al.}  [LHCb Collaboration],
  %``Test of lepton universality using $B^{+}\rightarrow K^{+}\ell^{+}\ell^{-}$ decays,''
  Phys.\ Rev.\ Lett.\  {\bf 113} (2014) 151601
  [arXiv:1406.6482 [hep-ex]].
  %%CITATION = ARXIV:1406.6482;%%
  %16 citations counted in INSPIRE as of 08 Dec 2014
  
\bibitem{P5prime}
R.~Aaij {\it et al.}  [LHCb Collaboration],
  %``Measurement of Form-Factor-Independent Observables in the Decay $B^{0} \to K^{*0} \mu^+ \mu^-$,''
  Phys.\ Rev.\ Lett.\  {\bf 111} (2013) 19,  191801
  [arXiv:1308.1707 [hep-ex]].
  %%CITATION = ARXIV:1308.1707;%%
  %80 citations counted in INSPIRE as of 08 Dec 2014
  
\bibitem{dimuon}
V.~M.~Abazov {\it et al.}  [D0 Collaboration],
  %``Evidence for an anomalous like-sign dimuon charge asymmetry,''
  Phys.\ Rev.\ Lett.\  {\bf 105} (2010) 081801
  [arXiv:1007.0395 [hep-ex]].
  %%CITATION = ARXIV:1007.0395;%%
  %148 citations counted in INSPIRE as of 08 Dec 2014
  
\bibitem{ttbarAFB}
T.~Aaltonen {\it et al.}  [CDF Collaboration],
  %``Measurement of the top quark forward-backward production asymmetry and its dependence on event kinematic properties,''
  Phys.\ Rev.\ D {\bf 87} (2013) 9,  092002
  [arXiv:1211.1003 [hep-ex]].
  %%CITATION = ARXIV:1211.1003;%%
  %117 citations counted in INSPIRE as of 08 Dec 2014
  
\bibitem{BEI}
G.~Blankenburg, J.~Ellis and G.~Isidori,
  %``Flavour-Changing Decays of a 125 GeV Higgs-like Particle,''
  Phys.\ Lett.\ B {\bf 712} (2012) 386
  [arXiv:1202.5704 [hep-ph]].
  %%CITATION = ARXIV:1202.5704;%%
  %49 citations counted in INSPIRE as of 08 Dec 2014
  
\bibitem{CMSHmutau}
CMS Collaboration, {\tt https://cds.cern.ch/record/1740976?ln=en}.

\bibitem{EHST}
J.~Ellis, D.~S.~Hwang, K.~Sakurai and M.~Takeuchi,
  %``Disentangling Higgs-Top Couplings in Associated Production,''
  JHEP {\bf 1404} (2014) 004
  [arXiv:1312.5736 [hep-ph]].
  %%CITATION = ARXIV:1312.5736;%%
  %22 citations counted in INSPIRE as of 08 Dec 2014
  
\bibitem{Bond}
J.~Bond {\it et al.}, {\tt http://www.imdb.com/title/tt0143145/}.

\bibitem{Buttazzo}
D.~Buttazzo, G.~Degrassi, P.~P.~Giardino, G.~F.~Giudice, F.~Sala, A.~Salvio and A.~Strumia,
  %``Investigating the near-criticality of the Higgs boson,''
  JHEP {\bf 1312} (2013) 089
  [arXiv:1307.3536].
  %%CITATION = ARXIV:1307.3536;%%
  %146 citations counted in INSPIRE as of 22 Aug 2014
  
\bibitem{D0mt}
A.~Jung, on behalf of the D0 Collaboration,
{\tt https://indico.cern.ch/event/279518/session/27/contribution/36/} {\tt material/slides/0.pdf.}

\bibitem{CMSmt}
CMS Collaboration, {\tt http://cds.cern.ch/record/1951019/files/TOP-14-015-pas.pdf}.

\bibitem{Moch}
S.~Moch, S.~Weinzierl, S.~Alekhin, J.~Blumlein, L.~de la Cruz, S.~Dittmaier, M.~Dowling and J.~Erler {\it et al.},
  %``High precision fundamental constants at the TeV scale,''
  arXiv:1405.4781 [hep-ph].
  %%CITATION = ARXIV:1405.4781;%%
  %26 citations counted in INSPIRE as of 08 Dec 2014
  
\bibitem{CMB}
P.~A.~R.~Ade {\it et al.}  [Planck Collaboration],
  %``Planck 2013 results. XXII. Constraints on inflation,''
  Astron.\ Astrophys.\  {\bf 571} (2014) A22
  [arXiv:1303.5082 [astro-ph.CO]];
  %%CITATION = ARXIV:1303.5082;%%
  %879 citations counted in INSPIRE as of 08 Dec 2014
  P.~A.~R.~Ade {\it et al.}  [BICEP2 Collaboration],
  %``Detection of B-Mode Polarization at Degree Angular Scales by BICEP2,''
  Phys.\ Rev.\ Lett.\  {\bf 112} (2014) 241101
  [arXiv:1403.3985 [astro-ph.CO]].
  %%CITATION = ARXIV:1403.3985;%%
  %860 citations counted in INSPIRE as of 08 Dec 2014
  
\bibitem{Oops}
M.~Fairbairn and R.~Hogan,
  %``Electroweak Vacuum Stability in light of BICEP2,''
  Phys.\ Rev.\ Lett.\  {\bf 112} (2014) 201801
  [arXiv:1403.6786 [hep-ph]];
  %%CITATION = ARXIV:1403.6786;%%
  %14 citations counted in INSPIRE as of 25 Aug 2014
A.~Hook, J.~Kearney, B.~Shakya and K.~M.~Zurek,
  %``Probable or Improbable Universe? Correlating Electroweak Vacuum Instability with the Scale of Inflation,''
  arXiv:1404.5953 [hep-ph].
  %%CITATION = ARXIV:1404.5953;%%
  %7 citations counted in INSPIRE as of 22 Aug 2014
 
 \bibitem{Sher}
 V.~Branchina, E.~Messina and M.~Sher,
  %``The lifetime of the electroweak vacuum and sensitivity to Planck scale physics,''
  arXiv:1408.5302 [hep-ph].
  %%CITATION = ARXIV:1408.5302;%%
  
\bibitem{SUSYmH}
J.~R.~Ellis, G.~Ridolfi and F.~Zwirner,
  %``Radiative corrections to the masses of supersymmetric Higgs bosons,''
  Phys.\ Lett.\ B {\bf 257} (1991) 83;
  %%CITATION = PHLTA,B257,83;%%
  %1239 citations counted in INSPIRE as of 08 Dec 2014
  H.~E.~Haber and R.~Hempfling,
  %``Can the mass of the lightest Higgs boson of the minimal supersymmetric model be larger than m(Z)?,''
  Phys.\ Rev.\ Lett.\  {\bf 66} (1991) 1815;
  %%CITATION = PRLTA,66,1815;%%
  %1232 citations counted in INSPIRE as of 08 Dec 2014
  Y.~Okada, M.~Yamaguchi and T.~Yanagida,
  %``Upper bound of the lightest Higgs boson mass in the minimal supersymmetric standard model,''
  Prog.\ Theor.\ Phys.\  {\bf 85} (1991) 1.
  %%CITATION = PTPKA,85,1;%%
  %1130 citations counted in INSPIRE as of 08 Dec 2014
  
\bibitem{FH}
T.~Hahn, S.~Heinemeyer, W.~Hollik, H.~Rzehak and G.~Weiglein,
  %``High-Precision Predictions for the Light CP -Even Higgs Boson Mass of the Minimal Supersymmetric Standard Model,''
  Phys.\ Rev.\ Lett.\  {\bf 112} (2014) 14,  141801
  [arXiv:1312.4937 [hep-ph]].
  %%CITATION = ARXIV:1312.4937;%%
  %53 citations counted in INSPIRE as of 08 Dec 2014
  
\bibitem{EHOW}
J.~R.~Ellis, S.~Heinemeyer, K.~A.~Olive and G.~Weiglein,
  %``Observability of the lightest CMSSM Higgs boson at hadron colliders,''
  Phys.\ Lett.\ B {\bf 515} (2001) 348
  [hep-ph/0105061].
  %%CITATION = HEP-PH/0105061;%%
  %48 citations counted in INSPIRE as of 08 Dec 2014
  
\bibitem{TLEP}
M.~Bicer {\it et al.}  [TLEP Design Study Working Group Collaboration],
  %``First Look at the Physics Case of TLEP,''
  JHEP {\bf 1401} (2014) 164
  [arXiv:1308.6176 [hep-ex]].
  %%CITATION = ARXIV:1308.6176;%%
  %48 citations counted in INSPIRE as of 23 Aug 2014
  
\bibitem{MC10}
O.~Buchmueller, R.~Cavanaugh, M.~Citron, A.~De Roeck, M.~J.~Dolan, J.~R.~Ellis, H.~Flaecher and S.~Heinemeyer {\it et al.},
  %``The NUHM2 after LHC Run 1,''
  arXiv:1408.4060 [hep-ph].
  %%CITATION = ARXIV:1408.4060;%%
  %1 citations counted in INSPIRE as of 23 Aug 2014
  
\bibitem{MC11}
O.~Buchmueller, R.~Cavanaugh, M.~Citron, A.~De Roeck, M.~J.~Dolan, J.~R.~Ellis, H.~Flaecher and S.~Heinemeyer {\it et al.},
{\it The pMSSM after LHC Run 1}, in preparation.
  
 \bibitem{futureg-2}
 FNAL g-2 Collaboration, {\tt http://muon-g-2.fnal.gov}; 
  H.~Iinuma (for the J-PARC New g-2/EDM experiment collaboration),
  {\tt http://iopscience.iop.org/1742-6596/295/1/012032/pdf/} {\tt 1742-6596\_295\_1\_012032.pdf}.

\bibitem{highWW}
K.~Rolbiecki and K.~Sakurai,
  %``Light stops emerging in WW cross section measurements?,''
  JHEP {\bf 1309} (2013) 004
  [arXiv:1303.5696 [hep-ph]];
  %%CITATION = ARXIV:1303.5696;%%
  %12 citations counted in INSPIRE as of 22 Aug 2014
  D.~Curtin, P.~Meade and P.~J.~Tien,
  %``Natural SUSY in Plain Sight,''
  arXiv:1406.0848 [hep-ph];
  %%CITATION = ARXIV:1406.0848;%%
  %6 citations counted in INSPIRE as of 22 Aug 2014
  J.~S.~Kim, K.~Rolbiecki, K.~Sakurai and J.~Tattersall,
  %```Stop' that ambulance! New physics at the LHC?,''
  arXiv:1406.0858 [hep-ph].
  %%CITATION = ARXIV:1406.0858;%%
  %6 citations counted in INSPIRE as of 22 Aug 2014
  
  \bibitem{CMSedge}
   CMS Collaboration, {\tt http://cds.cern.ch/record/1751493/files/SUS-12-019-pas.pdf}.
   
 \bibitem{WWNNLO}
 T.~Gehrmann, M.~Grazzini, S.~Kallweit, P.~Maierh\"ofer, A.~von Manteuffel, S.~Pozzorini, D.~Rathlev and L.~Tancredi,
  %``$W^+W^-$ production at hadron colliders in NNLO QCD,''
  arXiv:1408.5243 [hep-ph].
  %%CITATION = ARXIV:1408.5243;%%
  
  \bibitem{Economist}
  {\tt http://www.economist.com/blogs/graphicdetail/2012/07/daily-chart-1}.

\end{thebibliography}
\end{document}